# Development of a Real-Time Simulator Using EMTP-ATP Foreign models for Testing Relays

## September 17-18, 2018 - Arnhem, the Netherlands

**Renzo G. Fabián Espinoza**
**Independent, Peru/Brazil**

renzo.fe@ieee.org

**Lise R. Romero Navarrete**
**Allset Perú, Peru**

lrommel@allsetperu.com

***Abstract –*** *This paper reports the PC implementation of a real-time simulator for testing protective relays, based on the widely used EMTP-ATP software. The proposed simulator was implemented using the GNU/Linux OS with a real-time kernel. In order to generate the waveforms corresponding to simulated voltages and currents, a PCI card was used. This card also includes digital I/O interface. Via foreign models programmed in standard C, ATP was recompiled to include waveform generation at each simulation time step and digital I/O. Additionally, an IEC-61850 open source library was used, in order to use Sampled Values and GOOSE protocols. The resulting tool is a real-time simulator that can interact with protective relays by means of HiL tests. The performance of the simulator was analysed via an interaction with an actual relay.*

## 1 Introduction

The time-domain digital simulation of electromagnetic transients in power systems has evolved into the current real-time simulators, which can interact with real equipment via hardware-in-the-loop (HiL) digital real-time simulation (DRTS) [1]. These simulators have I/O interfaces that are used to interact with real devices and require high processing capacity; hence, they were initially implemented using parallel processors distributed in racks to test relays [2]. Currently, these simulators are very expensive.

EMTP-ATP, referred to here simply as ATP, is a non-commercial digital simulation program for electromagnetic transients used in electrical power systems and is based on royalty-free development of EMTP involving sophisticated models of power system components and control elements [3, 4, 5]. ATP provides the option to program new electrical and control components via MODELS that interact with ATP, receiving ATP-generated data as input and generating outputs that ATP receives and then uses in TACS/MODELS-controlled components. These components can also be programmed in non-MODELS languages (called foreign models), such as Fortran, C or C++ [6], enabling the use of various libraries, including hardware libraries, and integrating them in a new ATP executable file.

The authors of [7] report the development of an ATP-based real-time simulator that uses the Sampled Values and Goose IEC-61850 protocols to exchange simulated data via a network card interface. In [11] is reported a PC implementation of a real-time simulator using the sound card in order to generate analog outputs. It is limited to two channels and has latency problems that compromises the performance for this kind of application.

This work presents the development of a real-time simulator called here as Electromagnetic Transient Signal Exchanger in Real-Time (EMTSE-RT) or simply EMTSE.
In the EMTSE project is used an appropriate PCI card with 6 analog outputs and 16 digital I/O and implemented a specific real-time module to obtain a real-time simulator that can generate electrical waveforms to conduct HiL tests on protective relays. In this work, its integration with ATP is presented, but it can be integrated with other electric circuit solvers such as GNUCAP [10], for example.

## 2 Implementation

The EMTSE-RT was implemented in a PC with quad-core Intel-Xeon processor and 8 GB of RAM, using GNU/Linux OS with a real-time kernel (PREEMPT-RT), which enables the assignment of priorities to tasks and has preemption characteristics. ATP 64-bit Linux version is used and foreign models programmed in C are used to integrate ATP with the PCI and Ethernet card drivers according Figure 1. In each simulation time step, the numerical values of the electrical variables (such as voltages or currents) are externalized and the digital inputs and outputs (DIO) are read and write.

### 2.1 Testing via electrical interface

For this kind of tests, the multifunction PCI card is used. So, the electrical variables are externalized by means of 12-bit six-channel digital-to-analog converters (DACs) and injected to the protective relay under test by means of amplifiers. The digital outputs of the simulator are injected to the protective relay by means of a solid state relay module. The simulator receives actuations from the protective relay by means of a dry contact module.

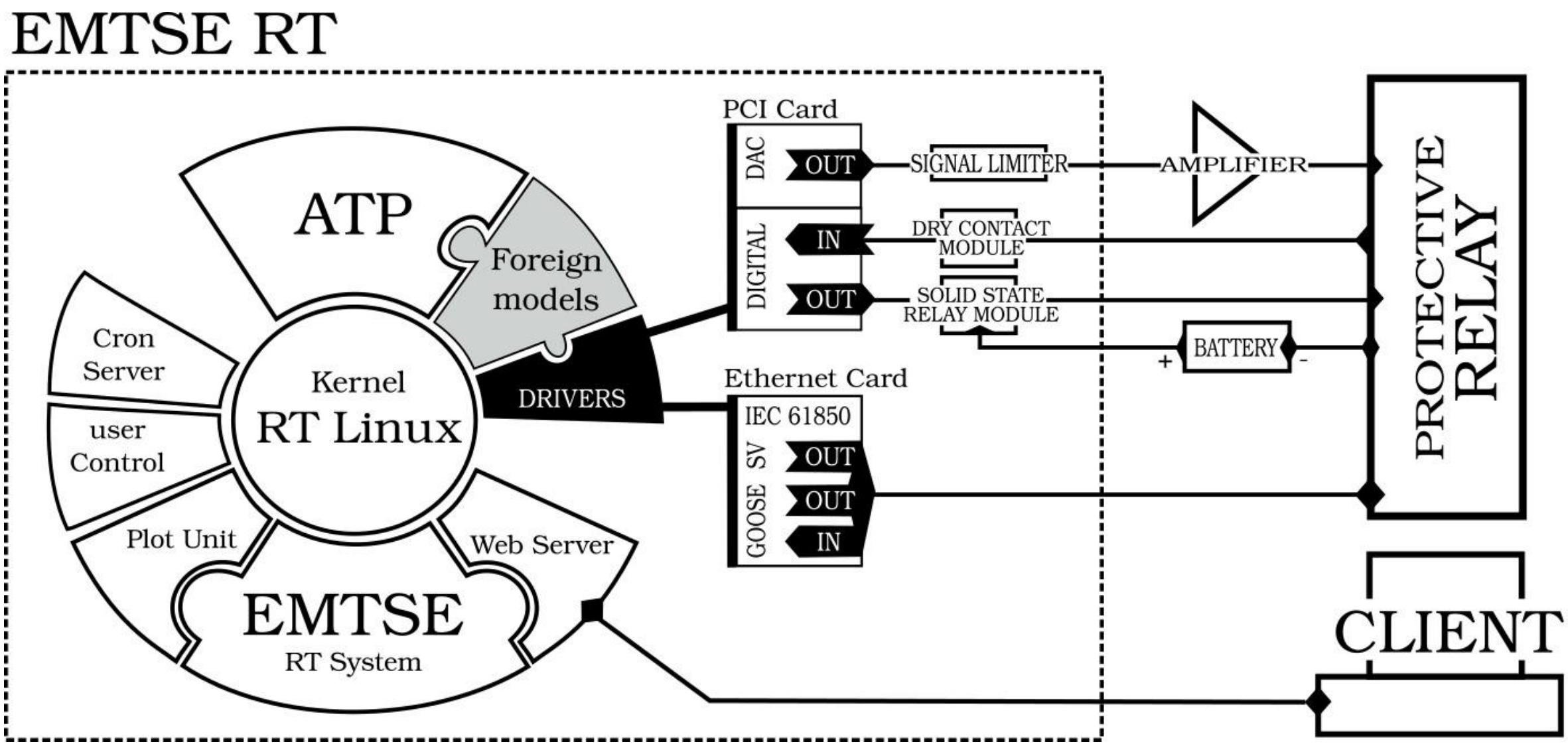

Figure 1 EMTSE RT diagram block

## 2.2 Testing via IEC 61850 Interface

In order to implement IEC 61850 protocols by means of foreign models, the open source library libIEC61850 was used. The project libIEC61850 provides a server and client library for IEC 61850 MMS, GOOSE and Sampled Values communication protocols written in C available under GPLv3 license [8].

## 2.3 Software interface

A web interface called Guépard tester was developed in order to conduct tests remotely. Figure 2 shows the login screen.

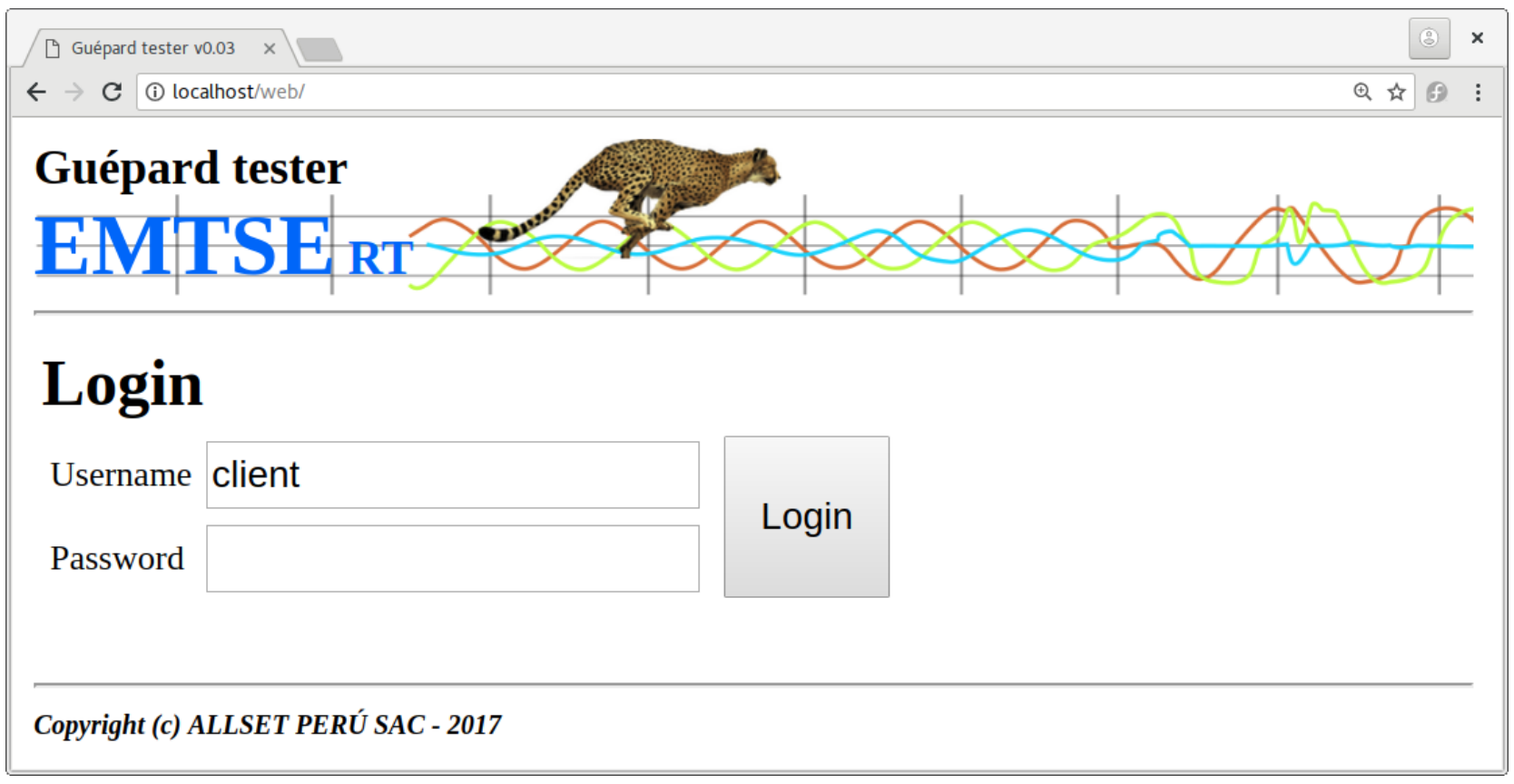

Figure 2 Login screen of Guépard tester

The interface allows the user to put in queu multiple ATP cases that will run in the target computer. This task in performed by means of a Cron server. Additionally, for the user, there is plot tool that allows monitoring some variables of interest. The user can choose the desired variables to plot. The list is automatically generated from pl4 file.

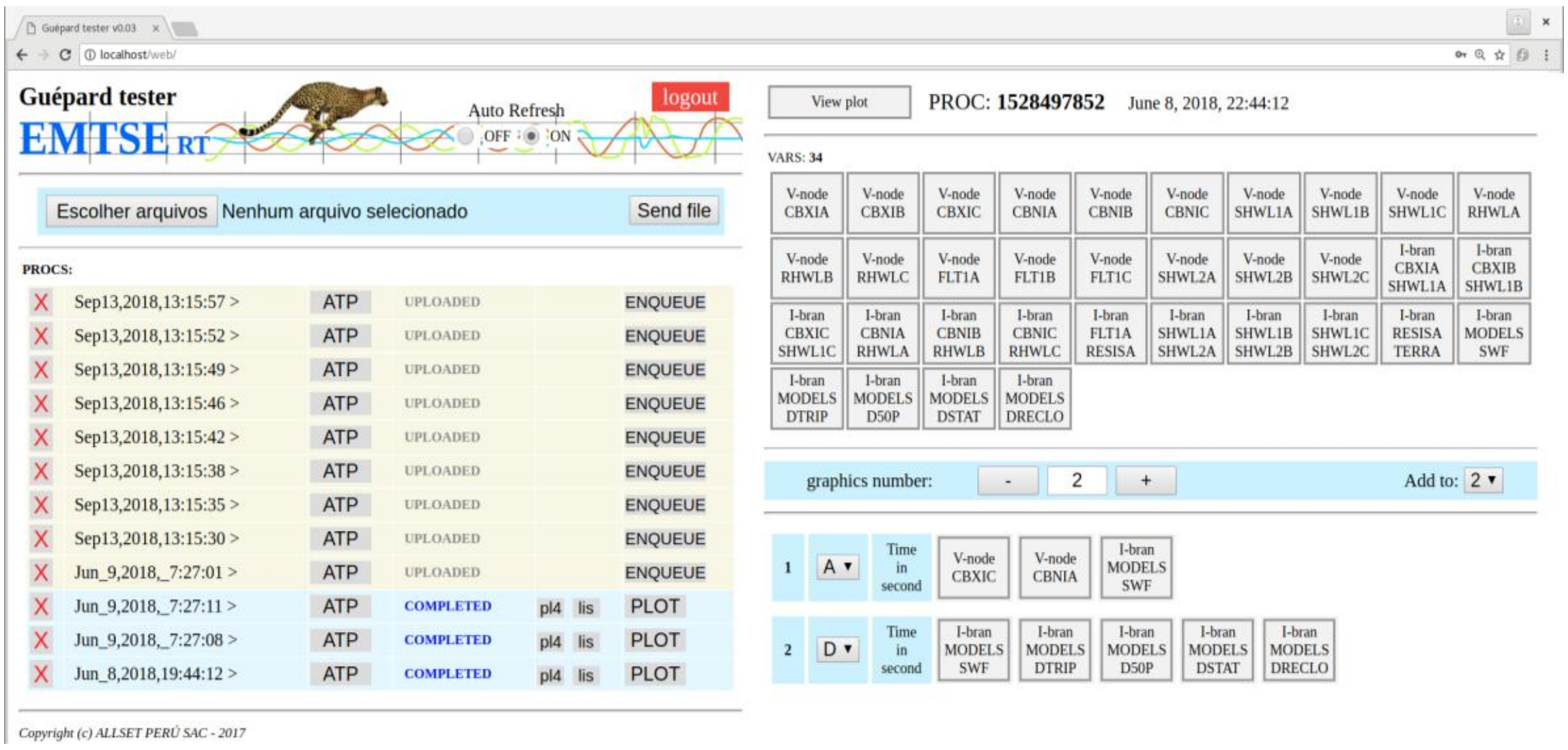

Figure 3 Guépard tester queue and plot tools

## 3 Test case

For testing, a 230 kV transmission system studied in [9] was considered. The transmission line is 180 km long, with the terminal systems being represented by their Thevenin equivalents, as shown in Figure 4. There is a CB between Bus 1 and the beginning of the transmission line.

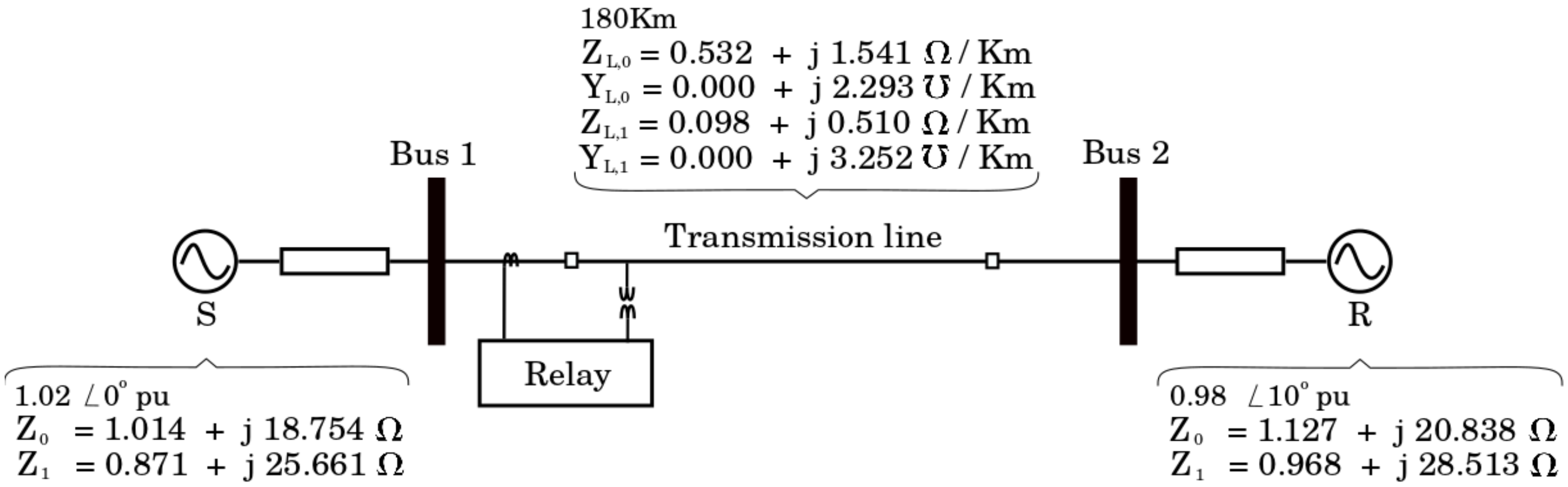

(a)

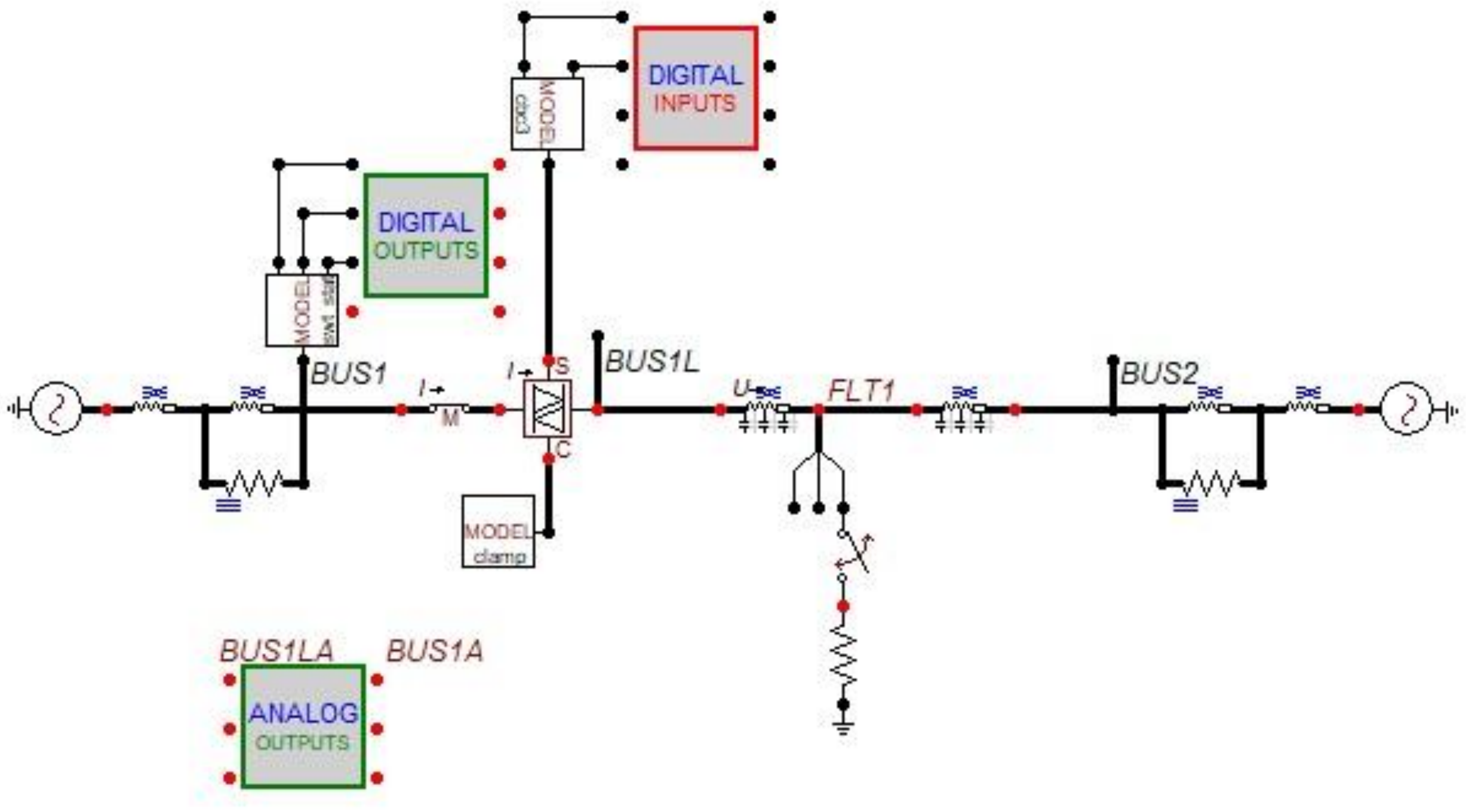

(b)

Figure 4 230 kV transmission system: (a) Single line diagram. (b) ATP-Draw schematic.

A HiL system was used to analyse the performance of EMTSE-RT by testing an actual relay. The analog outputs of the simulator were connected to an amplifier to obtain three voltages on the 115 V nominal voltage and three currents on the 5 A nominal current of the relay. Then, they were injected into the protective relay. The voltage and current undergo conversions of 2000:1 and 1200:5, respectively.

A schematic diagram of the hardware connection is shown in Figure 5(a).

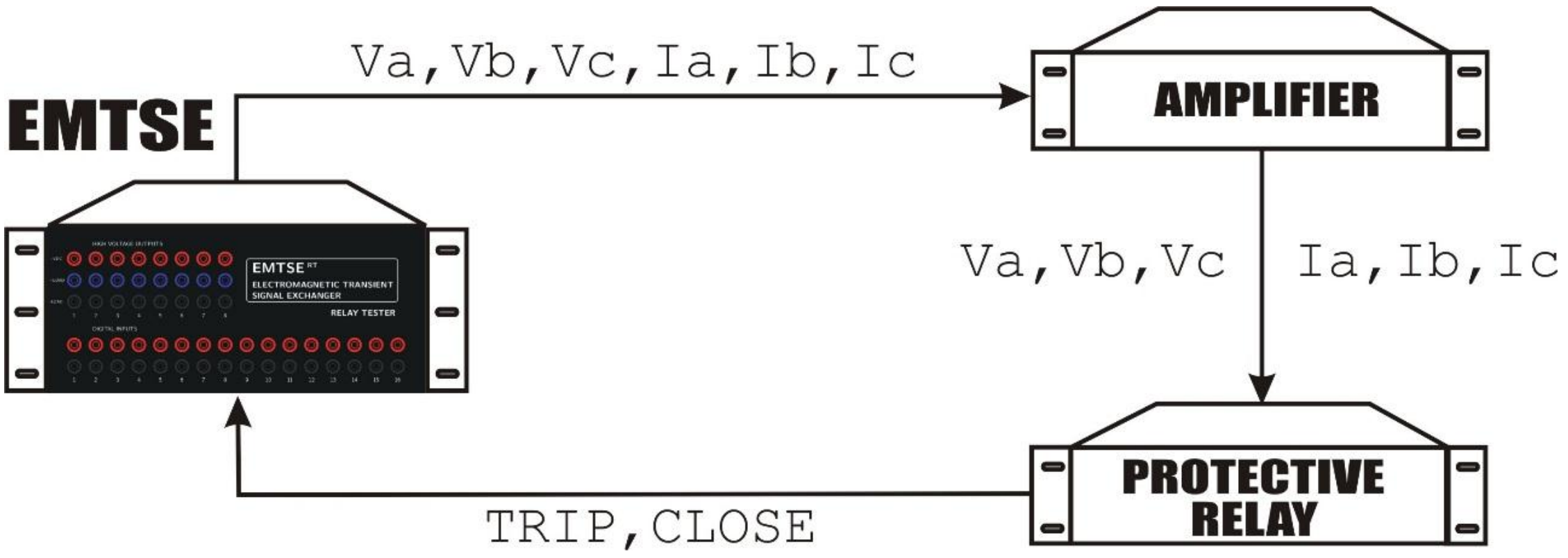

(a)

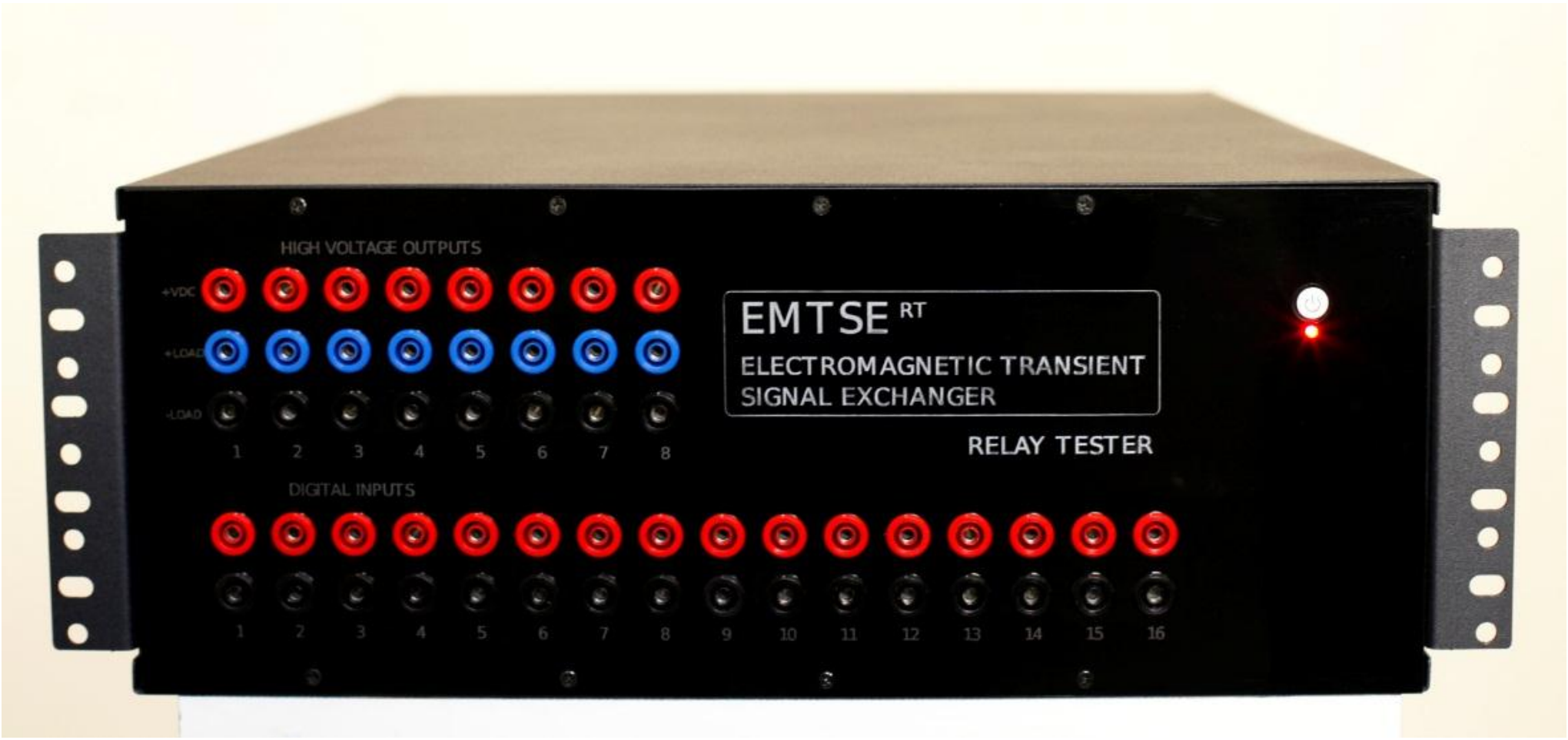

(b)

Figure 5 Laboratory hardware used for HiL tests. In (a) Schematic connection of all hardware. And in (b) Photography of EMTSE-RT prototype, courtesy of EMTSE project.

**4 Results**

Figure 6 shows the waveforms and some of the variables obtained in the ATP output file (.pl4) after simulation, where an AG fault of 50 ms time during, located 60 km from Bus 1 and with fault resistance of 9 Ohm was simulated. The relay operates after the fault, and EMTSE receives the trip signal to perform three-pole opening of the CB.

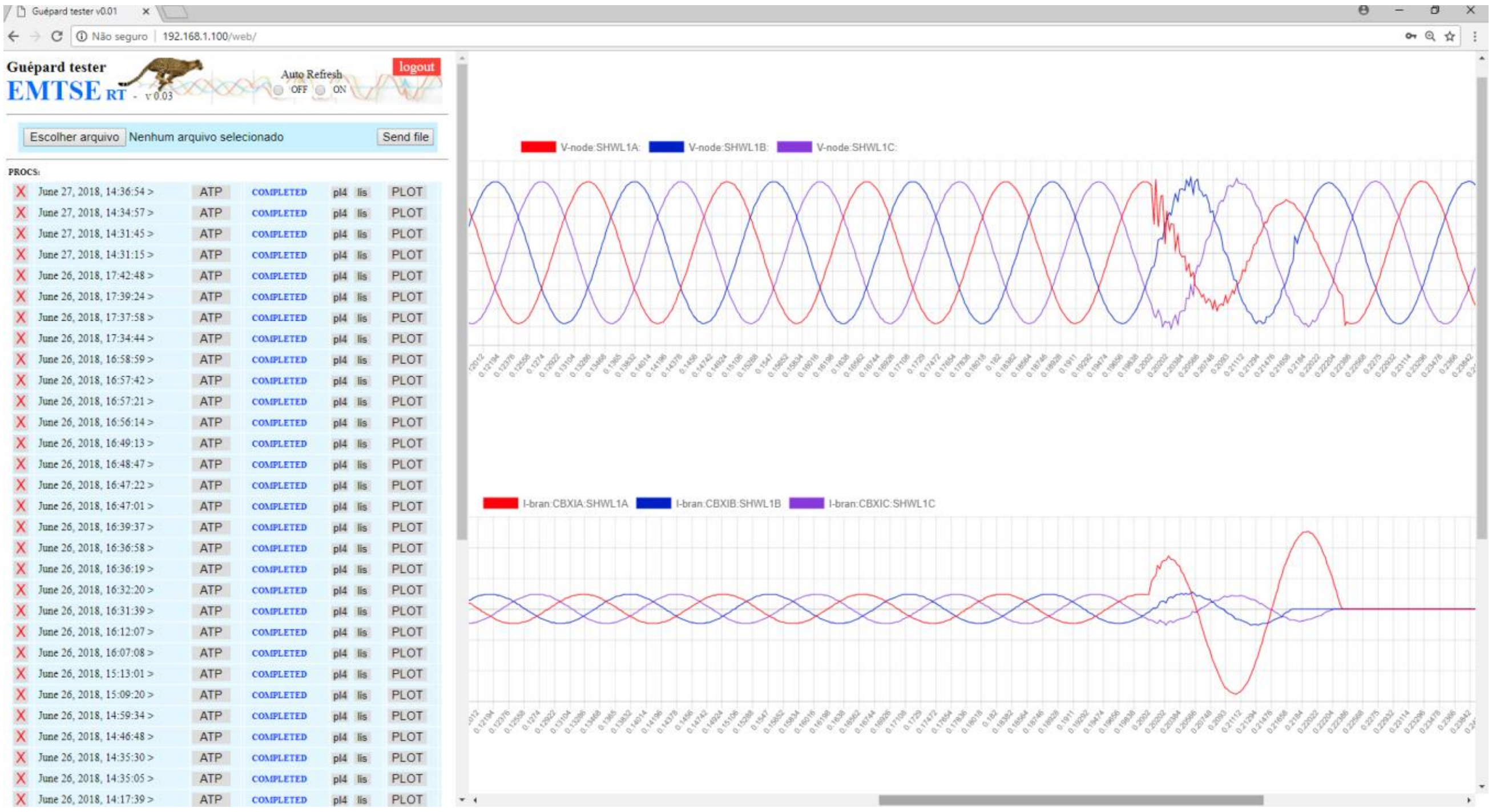

Figure 6 ATP simulated voltages and currents.

Figure 7 shows the current and voltage recorded by the relay (in COMTRADE format) for this case, including digital variables. It is possible to see that the waveforms recorded by the relay corresponds to the simulated by ATP. Furthermore, it is possible to see that the time of the signal trip to arrives to the relay is acceptable for this kind of tests.

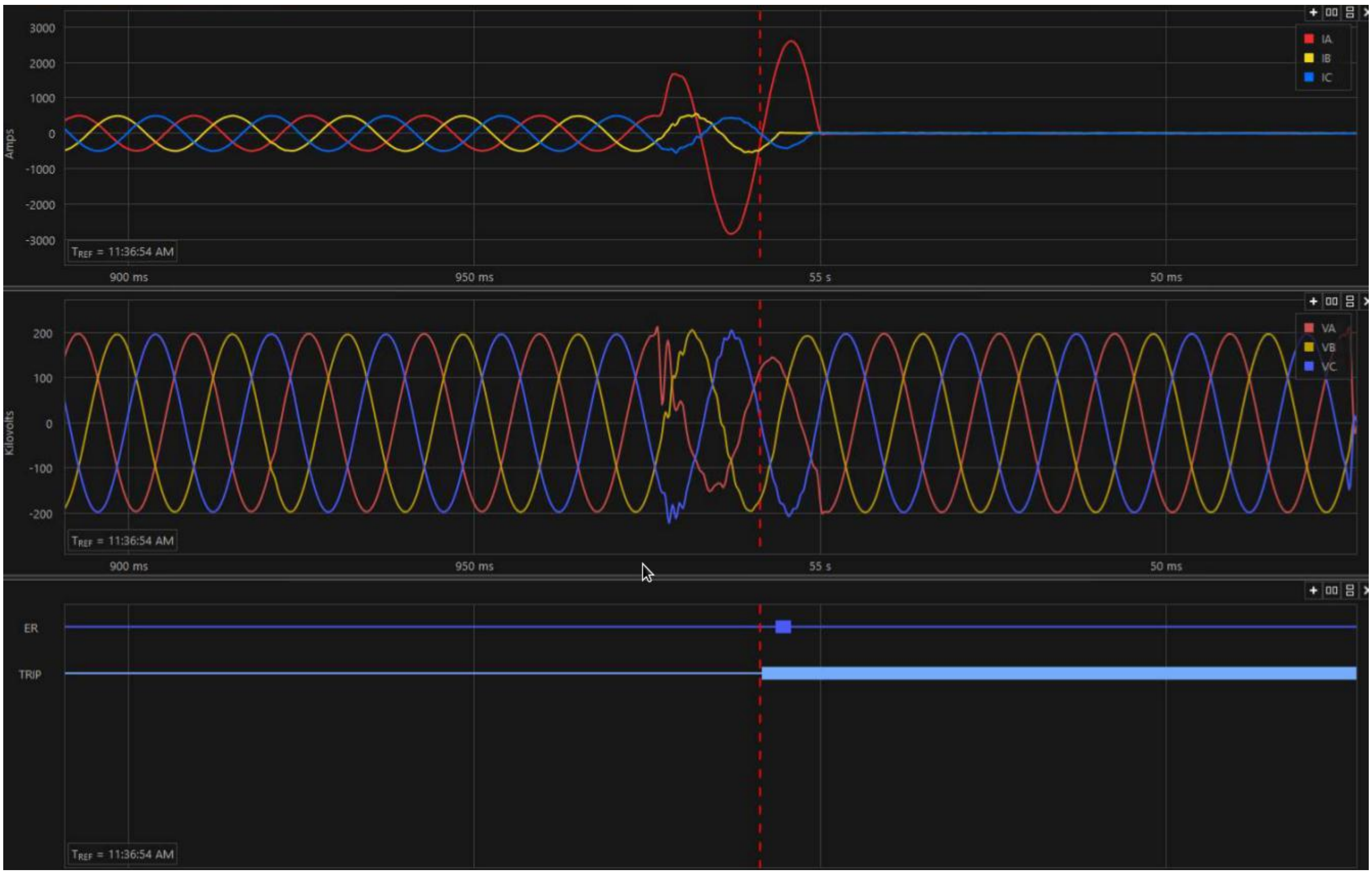

Figure 7 COMTRADE file recorded by the protective relay.

**5 Conclusions**

This paper presented the implementation of an ATP-based real-time simulator called EMTSE RT, which was used to conduct HiL testing of protective relays on a PC using a GNU/Linux with a real-time kernel via ATP foreign models programmed in standard C. A PCI with six analog output channels, eight digital inputs and eight digital outputs was used. The same principle can be used to implement IEC 61850 interface. A web interface was implemented in order to possibilite remote simulation and monitoring of the interest variables. This study demonstrated the possibility of conducting real-time HiL tests for a commercial relays using the developed EMTSE. A test generating and receiving signals in real time were conducted, with a simulation step size of 52 us. It is possible to see that the latency level is adequate for this kind of tests. This work shows the possibility of developing a real-time simulator to conduct HiL tests using low-cost hardware available for multiple vendors.

**References**

[1] M. D. Omar Faruque et al.: Real-Time Simulation Technologies for Power Systems Design, Testing, and Analysis, IEEE Power and Energy Technology Systems Journal, 2015, no 2, pp. 63-73.

[2] P. G. McLaren, R. Kuffel, R. Wierckx, J. Giesbrecht and L. Arendt: A real time digital simulator for testing relays, IEEE Transactions on Power Delivery, 1992, no. 1, pp. 207-213.

[3] Dommel, H.W.: Electromagnetic Transients Program (EMTP) TheoryBook, Portland OR: Bonneville Power Administration, 1986.

[4] CEU Group: Electromagnetic Transients Program (EMTP) RuleBook, 2001.

[5] Kizilcay, M.; Hoidalen, H. K.: EMTP-ATP, Numerical Analysis of Power System Transients and Dynamics, Chap. 2, Energy Engineering, 2015.

[6] Dube, L. : Models in ATP language manual, 1996.

[7] Ernesto Perez, Jaime de la Ree: Development of a real time simulator based on ATP-EMTP and sampled values of IEC61850-9-2, International Journal of Electrical Power & Energy Systems, 2016, pp 594-600.

[8] libIEC61850: Open source libraries for IEC 61850. http://libiec61850.com/libiec61850/

[9] K. M. Silva, A. Neves and B. A. Souza: Distance protection using a novel phasor estimation algorithm based on wavelet transform, 2008 IEEE Power and Energy Society General Meeting - Conversion and Delivery of Electrical Energy in the 21st Century, Pittsburgh, PA, 2008, pp. 1-8.

[10] Gnucap: Gnucap is a general purpose circuit simulator. It performs nonlinear dc and transient analyses, fourier analysis, and ac analysis. https://www.gnu.org/software/gnucap/gnucap.html

[11] Renzo G. Fabián Espinoza, Yuri Molina and Maria Tavares: PC Implementation of a Real-Time Simulator Using ATP Foreign Models and a Sound Card, Energies 2018, 11(8), 2140.